\documentclass[a4paper]{article}
\usepackage{multirow}
\usepackage[cp1250]{inputenc}
\usepackage{indentfirst}
\usepackage{bbm}
\title{Geometric-Algebra Quantum-Like Algorithms: Simon's Algorithm}
\author{Tomasz Magulski\footnote{email: magul@rudy.mif.pg.gda.pl}, Łukasz Orłowski\footnote{email: shangri@rudy.mif.pg.gda.pl}\\
\small \begin{em}Department of Theoretical Physics and Quantum Informatics\end{em}\\
\small \begin{em}Gdańsk University of Technology, 80-952 Gdańsk, Poland\end{em}}
\date{}
\begin{document}
\maketitle
\begin{abstract}
This is continuation of the approach to performing quantum algorithms using geometric structures which was presented by D. Aerts and M. Czachor in \cite{JPh}. We solve the Simon's problem which, next to the Shor's alghorithm, is a representative of quantum hidden subgroup class. We also highlight some advantages resulting from the fact that no quantum mechanics is involved. 
\end{abstract}
\section{Problem}
Consider a function
\begin{equation}
f:\{0,1\}^n \to \{0,1\}^m \textrm{, where }m \geq n-1
\end{equation}
we know that $f$ is 1-to-1, or there exists $s$ such that
\begin{equation}
\exists_{s \in \{0,1\}^n \setminus \{0^{(n)}\}} \forall_{x \neq y} f(x) = f(y) \iff x = y \oplus s, \label{def}
\end{equation}
where $\oplus $ denotes componentwise addition mod 2, i.e. n-dimensional XOR.\\
The problem is to determine which of these conditions is satisfied by $f$, and, in second case, to find $s$.\\

\section{Quantum Solution}
Let us use two quantum registers, which consist of $n$ and $m$ qbits respectively. We start with $n+m$ $0$s.
\begin{equation}
\vert \phi_0 \rangle = \vert \underbrace{0 \ldots 0}_n \rangle \vert \underbrace{0 \ldots 0}_m \rangle
\end{equation}
Quantum algorithm solves the problem using 3 quantum gates. At first the Hadamard gate acts on the first register. Note that using n-dimensional qubit space we have tensor power of n Hadamard gates. 
\begin{equation}
\vert \phi_1 \rangle = (H_n \vert 0^{(n)}\rangle) \vert 0^{(m)} \rangle = \frac{1}{\sqrt{2^n}} \left ( \sum_{k=0}^{2^n-1} \vert k \rangle \right ) \vert 0^{(m)}\rangle
\end{equation}
Second gate's action depends on function $f$. The gate fills the second register with function $f$ values, using first register as a set of arguments.
\begin{equation}
\vert \phi_2 \rangle = U_f \vert \phi_1 \rangle = \frac{1}{\sqrt{2^n}} \sum_{k=0}^{2^n-1}\vert k, f(k) \rangle
\end{equation}
Finally the Hadamard gate acts on the first register again.
\begin{equation}
\vert \phi_3 \rangle = \frac{1}{\sqrt{2^n}} \left ( \sum_{k=0}^{2^n-1} (H_n \vert k \rangle) \vert f(k)\rangle \right ) = \frac{1}{2^n} \sum_{i,k=0}^{2^n-1} (-1)^{\sigma (i,k)} \vert i, f(k) \rangle
\end{equation}
where $\sigma$ is a dot product of binary representations of $i$ and $k$ over $\{0,1\}^n$ space, i.e.
\begin{equation}
\sigma (i,j) = \sum_{k=1}^{n} i_k^{(2)} j_k^{(2)}
\end{equation}
Now we measure the first register.\\
Remember that if $f$ has the mask $s$ then $f(x) = f(x \oplus s)$, which means that our amplitude is a superposition of two amplitudes generated from $\vert k, f(k)\rangle$ and $\vert k \oplus s, f(k) \rangle$ in $\vert \phi_2 \rangle$.
\begin{equation}
\alpha_{i,f(k)} = \frac{1}{2^n}\left ((-1)^{\sigma(i,k)}+(-1)^{\sigma(i, k\oplus s)} \right )
\end{equation}
Employing
\begin{equation}
\sigma(i, k\oplus s)=\sigma(i,k)+\sigma(i,s) \textrm{ (mod } 2)
\end{equation}
then
\begin{equation}
\alpha_{i,f(k)} = \frac{1}{2^n} (-1)^{\sigma(i,k)} \left ( 1+(-1)^{\sigma(i,s)}\right )
\end{equation}
Now if $\sigma(i,s) = 1 \textrm{ (mod } 2)$ then both the amplitude and probability of getting such a state is $0$. This implies that every result of measurement satisfies
\begin{equation}
i_0 s_0 + i_1 s_1 + \cdots + i_{n-1} s_{n-1} = 0 \textrm{ (mod } 2)\label{rowdoukl}
\end{equation}
So in order to determine whether the mask~$s$ exists the procedure has to be repeated until $n-1$~linearly independent states~$i$ are found.\\
Then the following system of equations has to be solved 
\begin{equation}
\left \{
\begin{array}{ccc}
i_0^{(1)} s_0^* + i_1^{(1)} s_1^* + \cdots + i_{n-1}^{(1)} s_{n-1}^*&=&0 \textrm{ (mod } 2)\\
i_0^{(2)} s_0^* + i_1^{(2)} s_1^* + \cdots + i_{n-1}^{(2)} s_{n-1}^*&=&0 \textrm{ (mod } 2)\\
\vdots&\vdots&\vdots\\
i_0^{(n-1)} s_0^* + i_1^{(n-1)} s_1^* + \cdots + i_{n-1}^{(n-1)} s_{n-1}^*&=&0 \textrm{ (mod } 2)\\
\end{array}
 \right .\label{uklad}
\end{equation}
There are two possible solutions of this system - $0^{(n)}$ and~$s^*$.\\

Now recalling the definition we know that $f$ is 1-to-1 or, there exists $s$ which satisfies (\ref{def}). In the second case the non-trivial solution of (\ref{uklad}) is the mask we are looking for. In the first one it is easy to show that $s^*$ is just a random sequence.
The easiest way to determine which of these conditions are satisfied by $f$ is to check if 
$f(x)=f(x\oplus s^*)$ where $x\in \{0,1\}^n$.  
 
Originally Simon's problem was defined and solved in \cite{Sim}.
\pagebreak
\section{GA formulation}\label{ga}
GA formulation of the problem is based on the binary parametrization \cite{BSC}. Consider \linebreak  \mbox{$(n+m)$-dimensional} space with 
orthonormal basis $\{e_1 \ldots e_{n+m}\}$ and its associated GA.
Initial state is analogical to $\vert \phi_0 \rangle$:
\begin{equation}
e_{\underbrace{0 \ldots 0}_{n+m}}
\end{equation}
Let us use a multivector
\begin{equation}
E_n = \sum_{A_1,A_2,\ldots,A_n=0}^{1} e_{A_1A_2\ldots A_n\underbrace{0\ldots0}_m}
\end{equation} 
then
\begin{equation}\label{en}
E_n e_{\underbrace{0\ldots0}_{n+m}}= \sum_{A_1,A_2,\ldots,A_n=0}^{1} e_{A_1A_2\ldots A_n\underbrace{0\ldots0}_m}
\end{equation}
This step is equivalent to the first Hadamard gate in quantum algorithm. 
$U_f$ is an operation which performs as follows
\begin{equation}\label{box}
U_f E_n e_{\underbrace{0\ldots0}_{n+m}} = \sum_{A_1, A_2,\ldots,A_n=0}^{1} e_{A_1A_2\ldots A_nf_1(A_1A_2\ldots A_n)\ldots f_m(A_1A_2\ldots A_n)}
\end{equation}
Note that ~$f_i(A_1A_2\ldots A_n)$ is the $i$-th function $f$ value.\\
Let us consider the reverse of $E_n$
\begin{equation}
F_n = \sum_{B_1,B_2,\ldots,B_n=0}^{1} e_{B_1B_2\ldots B_n\underbrace{0\ldots0}_m}^{\dagger}
\end{equation}
Employing \cite{BSC}
\begin{equation}
e_{X_1X_2\ldots X_n}e_{Y_1Y_2\ldots Y_n} = (-1)^{\sum_{i<j} Y_i X_j}e_{(X_1\ldots X_n)\oplus(Y_1\ldots Y_n)}
\end{equation}
and
\begin{equation}
e_{X_1X_2\cdots X_n}^{\dagger} e_{X_1X_2\cdots X_n}=1
\end{equation}
we find
\begin{equation}
e_{X_1X_2\ldots X_n}^{\dagger} = (-1)^{\sum_{i<j} X_i X_j} e_{X_1X_2\ldots X_n}
\end{equation}
Therefore
\begin{equation}
F_n = \sum_{B_1,B_2,\ldots B_n=0}^{1} (-1)^{\sum_{i<j}B_i B_j} e_{B_1B_2\ldots B_n\underbrace{0\ldots0}_m}
\end{equation}

\pagebreak
$F_n$ performs as follows
\begin{eqnarray}\label{fn}
F_nU_fE_ne_{\underbrace{0\ldots0}_{n+m}}&=&\sum_{B_1,B_2,\ldots,B_n=0}^{1} \left ((-1)^{\sum_{i<j} B_i B_j} e_{B_1B_2\ldots B_n\underbrace{0\ldots0}_m} \right . \nonumber\\ && \left . \sum_{A_1,A_2,\ldots,A_n=0}^{1} e_{A_1A_2\ldots A_nf_1(A_1A_2\ldots A_n)\ldots f_m(A_1A_2\ldots A_n)} \right )\\
&=&\sum_{A_1,\ldots, A_n,B_1\ldots ,B_n=0}^{1}(-1)^{\sum_{i<j} B_i B_j}\nonumber\\ & &e_{B_1\ldots B_n\underbrace{0\ldots0}_{m}} e_{A_1\ldots A_nf_1(A_1\ldots A_n)\ldots f_m(A_1\ldots A_n)}\\
&=&\sum_{A_1,\ldots,A_n,B_1,\ldots ,B_n=0}^{1} (-1)^{\sum_{i<j} B_iB_j+\sum_{i<j} A_iB_j} \nonumber\\
&&e_{(B_1\ldots B_n\underbrace{0\ldots0}_{m})\oplus(A_1\ldots A_nf_1(A_1\ldots A_n)\ldots f_m(A_1\ldots A_n))}\label{FEU}
\end{eqnarray}

Now let us focus on the case in which $f$ has the mask $s$.
Think how the values standing next to each blade of the state (\ref{FEU}) look like.
Note that for each $f$ value the inverse image of $f$ is a two-element set. This implies that for all $(A_n)\in\{0,1\}^n$, in a state $U_fE_ne_{0\ldots0}$ there are exactly two blades with $f((A_n))$ values on the last $m$ bits:
\begin{equation}
e_{A_1\ldots A_nf_1(A_1\ldots A_n)\ldots f_m(A_1\ldots A_n)} \label{ostrzeaa}
\end{equation}
and
\begin{equation}
e_{(A_1\ldots A_n)\oplus(S_1\ldots S_n),f_1(A_1\ldots A_n)\ldots f_m(A_1\ldots A_n)}\label{ostrzeasa}
\end{equation}
where $(A_n)$ is arbitrary and $(S_n)$ is the mask we want to find.\\
Note that for all $(A_n)\in\{0,1\}^n$ there is a sequence in $\{0,1\}^n$ which by means of XOR operation can create an arbitrary sequence in $\{0,1\}^n$. Therefore, since $F_n$ contains all possible blades with $0$s on the last $m$ bits, its action on $U_fE_ne_{0\ldots0}$ gives us (from blades (\ref{ostrzeaa}) and (\ref{ostrzeasa})) two components of coefficient standing next to the blade:     
\begin{equation}
e_{X_1\ldots X_nf_1(A_1\ldots A_n)\ldots f_m(A_1\ldots A_n)}\label{ostrzexa}
\end{equation}
where $(X_n)$ is arbitrary.\\
Let us now consider $(X_n)$ and $(A_n)$. What is the coefficient standing next to (\ref{ostrzexa})?
It would be convenient to denote it by~$\alpha_{X,f(A)}$.\\
The first component resulting from (\ref{ostrzeaa}) is
\begin{equation}
(-1)^{\sum_{i<j} B_iB_j+\sum_{i<j}A_iB_j}=(-1)^{\sum_{i<j} (B_i+A_i)B_j}\label{amp1}
\end{equation}
where $(B_n)$ satisfies $(X_n)=(A_n)\oplus(B_n)$. The second component resulting from (\ref{ostrzeasa}) is
\begin{equation}
(-1)^{\sum_{i<j} B_i^{\oplus} B_j^{\oplus}+\sum_{i<j}(A_i\oplus S_i)B_j^{\oplus}}=(-1)^{\sum_{i<j} (B_i^{\oplus}+(A_i\oplus S_i))B_j^{\oplus}}\label{amp2}
\end{equation}
where $(B_n^{\oplus})$ satisfies $(X_n)=((A_n)\oplus(S_n))\oplus(B_n^{\oplus})$. Adding the values from (\ref{amp1}) and~(\ref{amp2}) we get the coefficient
\begin{equation}
\alpha_{X,f(A)}=(-1)^{\sum_{i<j} (B_i+A_i)B_j}+(-1)^{\sum_{i<j} (B_i^{\oplus}+(A_i\oplus S_i))B_j^{\oplus}}\label{amplituda1}
\end{equation}
which corresponds to a blade in a state $F_nU_fE_ne_{0\cdots 0}$.
The coefficients of this blade's binary parametrization carry no infromation about $(A_n)$ which means we also have no information about $(B_n)$, $(B_n^{\oplus})$, let alone $(S_n)$. However there are some conclusions to be drawn. Note that for all $i$
\begin{equation}
X_i = A_i \oplus B_i = (A_i \oplus S_i) \oplus B_i^{\oplus}\label{cotox}
\end{equation}
which implies
\begin{equation}
B_i^{\oplus}=A_i\oplus B_i\oplus A_i\oplus S_i
\end{equation}
\begin{equation}
B_i^{\oplus}=B_i\oplus S_i
\end{equation}
Employing this in (\ref{amplituda1}) we have
\begin{equation}
\alpha_{X,f(A)}=(-1)^{\sum_{i<j} \left (B_i+A_i\right) B_j}+(-1)^{\sum_{i<j}\left ( (B_i\oplus S_i)+(A_i\oplus S_i)\right) (B_j\oplus S_j)}
\end{equation}

$f(n)=(-1)^n$ is periodic with period 2 which allows us to switch some pluses for XOR operations and the other way
round
\begin{eqnarray}
\alpha_{X,f(A)}&=&(-1)^{\sum_{i<j} \left (B_i+A_i\right) B_j}+(-1)^{\sum_{i<j}\left (B_i\oplus S_i\oplus A_i\oplus S_i\right) (B_j+S_j)}\nonumber\\
&=&(-1)^{\sum_{i<j} \left (B_i+A_i\right) B_j}+(-1)^{\sum_{i<j}\left (B_i\oplus A_i\right) (B_j+S_j)}\nonumber\\
&=&(-1)^{\sum_{i<j} \left (B_i+A_i\right) B_j}+(-1)^{\sum_{i<j}\left (B_i+A_i\right) (B_j+S_j)}\nonumber\\
&=&(-1)^{\sum_{i<j} \left (B_i+A_i\right) B_j}+(-1)^{\sum_{i<j}\left (B_i+A_i\right)B_j+\sum_{i<j}\left (B_i+A_i\right) S_j}\nonumber\\
&=&(-1)^{\sum_{i<j} \left (B_i+A_i\right) B_j}\left (1+(-1)^{\sum_{i<j}\left (B_i+A_i\right) S_j}\right)\label{alfa}
\end{eqnarray}
Now we can see that whether~the particular blade occurs in the final multivector or zeros itself depends on the value of the exponent (expression in brackets). 

Note that unlike the quantum solution where each measurement gives us only one of the basic states from the superposition, in GA formulation the observation has no influence on the multivector so we can simply observe every particular amplitude of the blade we are interested in.

We have reached the point where we are able to determine whether $f$ is 1-to-1 or there 
exists the mask~$s$. According to (\ref{alfa}) $f$ has the mask if absolute values of amplitudes from the multivector $F_nU_fE_ne_{0\ldots 0}$ equal~2. On the other hand, if 
$f$ is 1-to-1, for all blades in (\ref{en}) the last m bits in (\ref{box}) are unique.
This implies that absolute values of all amplitudes in (\ref{fn}) equal 1.

Supposing that the mask $s$ exists let us focus on finding it.
From (\ref{alfa}) we have     
\begin{equation}
\left \{
\begin{array}{ll}
\sum_{i<j}\left (B_i+A_i\right) S_j = 0 \textrm{ (mod } 2) &\textrm{ for non-zero-blades}\\
\sum_{i<j}\left (B_i+A_i\right) S_j = 1 \textrm{ (mod } 2) &\textrm{ for zero-blades}
\end{array}
\right .
\end{equation}
and because of modulo 2 operation
\begin{equation}
\left \{
\begin{array}{ll}
\sum_{i<j}\left (B_i\oplus A_i\right) S_j = 0 \textrm{ (mod } 2) &\textrm{ for  non-zero-blades}\\
\sum_{i<j}\left (B_i\oplus A_i\right) S_j = 1 \textrm{ (mod } 2) &\textrm{ for zero-blades}
\end{array}
\right .
\end{equation}
Employing (\ref{cotox})
\begin{equation}
\left \{
\begin{array}{ll}
\sum_{i<j} X_i S_j = 0 \textrm{ (mod } 2) &\textrm{ for non-zero-blades}\\
\sum_{i<j} X_i S_j = 1 \textrm{ (mod } 2) &\textrm{ for zero-blades}
\end{array}
\right .
\end{equation}

\noindent
In terms of Simon's problem we examine the amplitudes of blades which are in the form of 
\begin{equation}
e_{\underbrace{0\ldots 0}_k 11 \underbrace{0\ldots 0}_{n+m-k-2}}\label{jedynki}
\end{equation}
where $k \in \langle 0, n-2 \rangle$.
Returning to our system we get
\begin{equation}
\left \{
\begin{array}{ll} 
\sum_{i<j<k+2} X_i S_j + \sum_{i<k+2} X_i S_{k+2} + \sum_{i<j, j>k+2} X_i S_j = 0 \textrm{ (mod } 2) &\textrm{ for non-zero-blades}\\
\sum_{i<j<k+2} X_i S_j + \sum_{i<k+2} X_i S_{k+2} + \sum_{i<j, j>k+2} X_i S_j = 1 \textrm{ (mod } 2) &\textrm{ for zero-blades}
\end{array}
\right . \label{}
\end{equation}
Employing (\ref{jedynki})
\begin{equation}
\left \{
\begin{array}{ll}
\sum_{1<j<k+2} 0 S_j + 1 S_{k+2} + \sum_{j>k+2} 2 S_j = 0 \textrm{ (mod } 2) &\textrm{ for non-zero-blades}\\
\sum_{1<j<k+2} 0 S_j + 1 S_{k+2} + \sum_{j>k+2} 2 S_j = 1 \textrm{ (mod } 2) &\textrm{ for zero-blades}
\end{array}
\right .
\end{equation}
Again because of the modulo 2 operation
\begin{equation}
\left \{
\begin{array}{ll}
S_{k+2}=0 \textrm{ (mod } 2) &\textrm{ for non-zero-blades}\\
S_{k+2}=1 \textrm{ (mod } 2) &\textrm{ for zero-blades}
\end{array}
\right .\label{wynik}
\end{equation}
Now we can determine all the bits of our mask from $S_2$ to $S_n$. To find $S_1$ we have to check two possible masks 
$s$: $(0S_2\cdots S_n)$ and $(1S_2\cdots S_n)$.

\section{Explicit Examples}\label{examples}

In this section we want to show explicitly how GA formulation works. We present two examples.

Let us consider the following 1-to-1 function
\begin{table}[h]
\begin{center}
\begin{tabular}{|c|cccc|}
\hline
\multirow{2}{*}{$X$} & 0&0&1&1\\
    & 0&1&0&1\\
\hline \hline
\multirow{2}{*}{$f(X)$} & 1&0&1&0\\
			 & 0&0&1&1\\
\hline
\end{tabular}
\caption{1-to-1 function}\label{f1to1}
\end{center}
\end{table}

and perform Simon's algorithm
\begin{equation}
e_{0000}
\end{equation}
\begin{equation}
E_n e_{0000}=e_{0000}+e_{0100}+e_{1000}+e_{1100}
\end{equation}
\begin{equation}
U_f E_n e_{0000}=e_{0010}+e_{0100}+e_{1011}+e_{1101}
\end{equation}
\begin{eqnarray}
F_n U_f E_n e_{0000}&=&e_{0000}+e_{0001}+e_{0010}+e_{0011}+e_{0100}+e_{0101}+e_{0110}+e_{0111}\nonumber \\
										& &-e_{1000}-e_{1001}+e_{1010}+e_{1011}+e_{1100}+e_{1101}-e_{1110}-e_{1111}\label{wynik121}
\end{eqnarray}
We can see that the absolute value of every amplitude in the multivector is 1 which proves that  $f$ is 1-to-1.

\pagebreak
Now let us define the function which has a mask

\begin{table}[h]
\begin{center}
\begin{tabular}{|c|cccccccc|}
\hline
\multirow{3}{*}{$X$} & 0&0&0&0&1&1&1&1\\
    & 0&0&1&1&0&0&1&1\\
    & 0&1&0&1&0&1&0&1\\
\hline \hline
\multirow{2}{*}{$f(X)$} & 1&0&1&0&1&0&1&0\\
			 & 1&0&1&0&0&1&0&1\\
\hline
\end{tabular}\caption{Function with mask $s=(010)$}\label{fmask}
\end{center}\end{table}

\noindent
and perform the algorithm
\begin{equation}
e_{00000}
\end{equation}
\begin{equation}
E_n e_{00000}=e_{00000}+e_{00100}+e_{01000}+e_{01100}+e_{10000}+e_{10100}+e_{11000}+e_{11100}
\end{equation}
\begin{equation}
U_f E_n e_{00000}=e_{00011}+e_{00100}+e_{01011}+e_{0110}+e_{10010}+e_{10101}+e_{11010}+e_{11100}
\end{equation}
\begin{eqnarray}
F_n U_f E_n e_{00000}&=&2(e_{00000}+e_{00001}+e_{00010}+e_{00011}+e_{00100}+e_{00101}+e_{00110}+e_{00111}\nonumber\\            & &-e_{01000}-e_{01001}+e_{01010}+e_{01011}+e_{01100}+e_{01101}-e_{01110}-e_{01111})
\end{eqnarray}
In this case the absolute value of every amplitude in the multivector equals 2 and therefore the non-trivial mask $s$ exists.

In Sec. \ref{ga} we showed how to determine our mask $s$ using blades in the form of (\ref{jedynki}).\\
Let us illustrate how it works with the help of the following table 
\begin{table}[h]
\begin{center}
\begin{tabular}{|c|cccccccc||cc|}
\hline
\multirow{5}{*}{blade's binary parametrization}& 0& 0& 0& 0&1&1&1&1&0&1\\
															 & 1&1&1&1&1&1&1&1&1&1\\
															 &1&1&1&1& 0& 0& 0& 0&1&0\\
															 & 0& 0&1&1& 0& 0&1&1&X&X\\
															 & 0&1& 0&1& 0&1& 0&1&X&X\\
\hline \hline
absolute value of an amplitude   & 2&2&2&2&0&0&0&0&2&0\\
\hline
\end{tabular}\caption{Blade's amplitudes}
\end{center}\end{table}

\noindent Thanks to (\ref{wynik}) we have\\
\begin{equation}
\left \{ 
\begin{array}{ll}
S_{0+2}=S_2=1 &\textrm{ because the amplitude of }e_{110XX}\textrm{ equals 0}\\
S_{1+2}=S_3=0 &\textrm{ because the amplitude of }e_{011XX}\textrm{ equals 2}\\
\end{array}
\right .
\end{equation}
and therefore $s=(X10)$. To determine the first bit we need to evaluate $f(000)$ and $f(010)$.
In our example $f(000)=f(010)$ so $S_1=0$. In this example the algorithm has proved that the mask $s$ exists and equals~$(010)$.  
\pagebreak

\section{Cartan's representation}
Let us use the matrix algebra known as Cartan's representation of 1-blade in GA \cite{BudTra}:
\begin{eqnarray}
e_{2k-1}&=&\underbrace{\sigma_1 \otimes \cdots \otimes \sigma_1}_{n-k}\otimes \sigma_3 \otimes \underbrace{\mathbbm{1} \otimes \cdots \otimes \mathbbm{1}}_{k-1}\label{cartan1}\\
e_{2k}&=&\underbrace{\sigma_1 \otimes \cdots \otimes \sigma_1}_{n-k}\otimes \sigma_2 \otimes \underbrace{\mathbbm{1} \otimes \cdots \otimes \mathbbm{1}}_{k-1}
\label{cartan2}
\end{eqnarray}\\
and obviously the scalar representation is 
\begin{equation}
e_0=\underbrace{\mathbbm{1}\otimes \cdots \otimes \mathbbm{1}}_{n}
\end{equation}
where $\sigma_1$, $\sigma_2$, $\sigma_3$ are the Pauli matrices and $\mathbbm{1}$ denotes $2\times 2$ unity matrix.\\
An arbitrary blade can be represented by the adequate product of (\ref{cartan1}) and (\ref{cartan2}).\\
We know that
\begin{equation} 
\textrm{Tr }\sigma_1=\textrm{Tr }\sigma_2=\textrm{Tr }\sigma_3=0
\end{equation}
and
\begin{equation}
\textrm{Tr } \left(\bigotimes_{i=1}^{n} A_i \right)=\prod_{i=1}^{n} \textrm{Tr }A_i
\end{equation}
\begin{equation}
\textrm{Tr } \left(\sum_{i=1}^{n} A_i \right)=\sum_{i=1}^{n} \textrm{Tr }A_i
\end{equation}
Therefore for the scalar
\begin{equation}
\textrm{Tr }e_0=\left(\textrm{Tr } \mathbbm{1}\right)^n=2^n
\end{equation}
and for 1-blades
\begin{equation}
\textrm{Tr } e_{2k-1}=\left(\textrm{Tr } \sigma_1\right)^{n-k}\textrm{Tr } \sigma_3 \left(\textrm{Tr } \mathbbm{1}\right)^{k-1}=0
\end{equation}
\begin{equation}
\textrm{Tr } e_{2k}=\left(\textrm{Tr } \sigma_1\right)^{n-k}\textrm{Tr } \sigma_2 \left(\textrm{Tr } \mathbbm{1}\right)^{k-1}=0
\end{equation}
It is easy to show that
\begin{equation}
\left \{
\begin{array}{ll}
\textrm{Tr }(e_{A_1\ldots A_n}e_{B_1\ldots B_n}) = 0 & \textrm{if $\exists_i A_i \neq B_i$}\\
\textrm{Tr }({e_{A_1\ldots A_n}}^2)=(-1)^{\frac{\sum_{i=1}^{n} A_i (\sum_{i=1}^{n} A_i-1)}{2}} 2^n&
\end{array}
\right .
\end{equation}
So having a multivector
\begin{equation}
X=\sum_{A_1,\ldots ,A_n=0}^{1} X_{A_1\ldots A_n} e_{A_1\ldots A_n}
\end{equation}
we can compute the coefficients by
\begin{equation}
X_{A_1\ldots A_n}=\frac{(-1)^{\frac{\sum_{i=1}^{n} A_i (\sum_{i=1}^{n} A_i-1)}{2}}}{2^n}\textrm{Tr } (e_{A_1\ldots A_n} X)\label{ca2mu}
\end{equation}

Let us consider the problem for the 1-to-1 function from the previous section (Table \ref{f1to1}).\\
The initial state $e_{0000}$ has the following representation:
\begin{equation}
\begin{tiny}
e_{0000}=
\left(
\begin{array}{cccccccccccccccc}
1&0&0&0&0&0&0&0&0&0&0&0&0&0&0&0\\
0&1&0&0&0&0&0&0&0&0&0&0&0&0&0&0\\
0&0&1&0&0&0&0&0&0&0&0&0&0&0&0&0\\
0&0&0&1&0&0&0&0&0&0&0&0&0&0&0&0\\
0&0&0&0&1&0&0&0&0&0&0&0&0&0&0&0\\
0&0&0&0&0&1&0&0&0&0&0&0&0&0&0&0\\
0&0&0&0&0&0&1&0&0&0&0&0&0&0&0&0\\
0&0&0&0&0&0&0&1&0&0&0&0&0&0&0&0\\
0&0&0&0&0&0&0&0&1&0&0&0&0&0&0&0\\
0&0&0&0&0&0&0&0&0&1&0&0&0&0&0&0\\
0&0&0&0&0&0&0&0&0&0&1&0&0&0&0&0\\
0&0&0&0&0&0&0&0&0&0&0&1&0&0&0&0\\
0&0&0&0&0&0&0&0&0&0&0&0&1&0&0&0\\
0&0&0&0&0&0&0&0&0&0&0&0&0&1&0&0\\
0&0&0&0&0&0&0&0&0&0&0&0&0&0&1&0\\
0&0&0&0&0&0&0&0&0&0&0&0&0&0&0&1\\
\end{array}\right)\end{tiny}\label{e0000}
\end{equation}
Note that
\begin{equation}
\begin{tiny}
E_n=E_n e_{0000}=
\left(
\begin{array}{cccccccccccccccc}
1&-i&0&0&0&0&0&0&0&0&0&0&0&0&1&-i\\
-i&1&0&0&0&0&0&0&0&0&0&0&0&0&i&-1\\
0&0&1&-i&0&0&0&0&0&0&0&0&1&-i&0&0\\
0&0&-i&1&0&0&0&0&0&0&0&0&i&-1&0&0\\
0&0&0&0&1&-i&0&0&0&0&1&-i&0&0&0&0\\
0&0&0&0&-i&1&0&0&0&0&i&-1&0&0&0&0\\
0&0&0&0&0&0&1&-i&1&-i&0&0&0&0&0&0\\
0&0&0&0&0&0&-i&1&i&-1&0&0&0&0&0&0\\
0&0&0&0&0&0&1&-i&1&-i&0&0&0&0&0&0\\
0&0&0&0&0&0&i&-1&-i&1&0&0&0&0&0&0\\
0&0&0&0&1&-i&0&0&0&0&1&-i&0&0&0&0\\
0&0&0&0&i&-1&0&0&0&0&-i&1&0&0&0&0\\
0&0&1&-i&0&0&0&0&0&0&0&0&1&-i&0&0\\
0&0&i&-1&0&0&0&0&0&0&0&0&-i&1&0&0\\
1&-i&0&0&0&0&0&0&0&0&0&0&0&0&1&-i\\
i&-1&0&0&0&0&0&0&0&0&0&0&0&0&-i&1\\
\end{array}\right)\end{tiny}\label{Ene0000}
\end{equation}
In general the second gate is not a multivector so it does not have Cartan's representation and  therefore only the result of $U_f E_n e_{0000}$ can be represented:
\begin{equation}
\begin{tiny}
\left(
\begin{array}{cccccccccccccccc}
0&0&0&0&0&0&0&0&0&0&0&0&1-i&0&0&-1-i\\
0&0&0&0&0&0&0&0&0&0&0&0&0&1+i&-1+i&0\\
0&0&0&0&0&0&0&0&0&0&0&0&0&1-i&-1-i&0\\
0&0&0&0&0&0&0&0&0&0&0&0&1+i&0&0&-1+i\\
0&0&0&0&0&0&0&0&1-i&0&0&-1-i&0&0&0&0\\
0&0&0&0&0&0&0&0&0&1+i&-1+i&0&0&0&0&0\\
0&0&0&0&0&0&0&0&0&1-i&-1-i&0&0&0&0&0\\
0&0&0&0&0&0&0&0&1+i&0&0&-1+i&0&0&0&0\\
0&0&0&0&1-i&0&0&-1-i&0&0&0&0&0&0&0&0\\
0&0&0&0&0&1+i&-1+i&0&0&0&0&0&0&0&0&0\\
0&0&0&0&0&1-i&-1-i&0&0&0&0&0&0&0&0&0\\
0&0&0&0&1+i&0&0&-1+i&0&0&0&0&0&0&0&0\\
1-i&0&0&-1-i&0&0&0&0&0&0&0&0&0&0&0&0\\
0&1+i&-1+i&0&0&0&0&0&0&0&0&0&0&0&0&0\\
0&1-i&-1-i&0&0&0&0&0&0&0&0&0&0&0&0&0\\
1+i&0&0&-1+i&0&0&0&0&0&0&0&0&0&0&0&0\\
\end{array}\right)\end{tiny}
\end{equation}
Then
\begin{equation}
\begin{tiny}
F_n=
\left(
\begin{array}{cccccccccccccccc}
1&i&0&0&0&0&0&0&0&0&0&0&0&0&1&-i\\
i&1&0&0&0&0&0&0&0&0&0&0&0&0&i&-1\\
0&0&1&i&0&0&0&0&0&0&0&0&1&-i&0&0\\
0&0&i&1&0&0&0&0&0&0&0&0&i&-1&0&0\\
0&0&0&0&1&i&0&0&0&0&1&-i&0&0&0&0\\
0&0&0&0&i&1&0&0&0&0&i&-1&0&0&0&0\\
0&0&0&0&0&0&1&i&1&-i&0&0&0&0&0&0\\
0&0&0&0&0&0&i&1&i&-1&0&0&0&0&0&0\\
0&0&0&0&0&0&1&-i&1&i&0&0&0&0&0&0\\
0&0&0&0&0&0&i&-1&i&1&0&0&0&0&0&0\\
0&0&0&0&1&-i&0&0&0&0&1&i&0&0&0&0\\
0&0&0&0&i&-1&0&0&0&0&i&1&0&0&0&0\\
0&0&1&-i&0&0&0&0&0&0&0&0&1&i&0&0\\
0&0&i&-1&0&0&0&0&0&0&0&0&i&1&0&0\\
1&-i&0&0&0&0&0&0&0&0&0&0&0&0&1&i\\
i&-1&0&0&0&0&0&0&0&0&0&0&0&0&i&1\\
\end{array} \right)\end{tiny}\label{Fn}
\end{equation}
and its action on $U_f E_n e_{0000}$
\begin{equation}
\begin{tiny}
\left(
\begin{array}{cccccccccccccccc}
1-i&1-i&-1-i&1+i&0&0&0&0&0&0&0&0&1-i&-1+i&-1-i&-1-i\\
-1-i&1+i&1-i&1-i&0&0&0&0&0&0&0&0&1+i&1+i&-1+i&1-i\\
1-i&1-i&1+i&-1-i&0&0&0&0&0&0&0&0&-1+i&1-i&-1-i&-1-i\\
1+i&-1-i&1-i&1-i&0&0&0&0&0&0&0&0&1+i&1+i&1-i&-1+i\\
0&0&0&0&1-i&1-i&-1-i&1+i&1-i&-1+i&-1-i&-1-i&0&0&0&0\\
0&0&0&0&-1-i&1+i&1-i&1-i&1+i&1+i&-1+i&1-i&0&0&0&0\\
0&0&0&0&1-i&1-i&1+i&-1-i&-1+i&1-i&-1-i&-1-i&0&0&0&0\\
0&0&0&0&1+i&-1-i&1-i&1-i&1+i&1+i&1-i&-1+i&0&0&0&0\\
0&0&0&0&1-i&-1+i&-1-i&-1-i&1-i&1-i&-1-i&1+i&0&0&0&0\\
0&0&0&0&1+i&1+i&-1+i&1-i&-1-i&1+i&1-i&1-i&0&0&0&0\\
0&0&0&0&-1+i&1-i&-1-i&-1-i&1-i&1-i&1+i&-1-i&0&0&0&0\\
0&0&0&0&1+i&1+i&1-i&-1+i&1+i&-1-i&1-i&1-i&0&0&0&0\\
1-i&-1+i&-1-i&-1-i&0&0&0&0&0&0&0&0&1-i&1-i&-1-i&1+i\\
1+i&1+i&-1+i&1-i&0&0&0&0&0&0&0&0&-1-i&1+i&1-i&1-i\\
-1+i&1-i&-1-i&-1-i&0&0&0&0&0&0&0&0&1-i&1-i&1+i&-1-i\\
1+i&1+i&1-i&-1+i&0&0&0&0&0&0&0&0&1+i&-1-i&1-i&1-i\\
\end{array} \right)\end{tiny}
\end{equation}
Now using (\ref{ca2mu}) we can find the amplitudes (Table \ref{trace1to1})

\begin{table}[h]
\begin{center}\begin{small}
\begin{tabular}{|c|cccccccccccccccc|}
\hline
$A_1$&0&0&0&0&0&0&0&0&1&1&1&1&1&1&1&1\\
$A_2$&0&0&0&0&1&1&1&1&0&0&0&0&1&1&1&1\\
$A_3$&0&0&1&1&0&0&1&1&0&0&1&1&0&0&1&1\\
$A_4$&0&1&0&1&0&1&0&1&0&1&0&1&0&1&0&1\\
\hline
\hline
Tr($e_{A_1A_2A_3A_4} F_n U_f E_n e_{0000}$)&16&16&16&-16&16&-16&-16&-16&-16&16&-16&-16&-16&-16&16&-16\\
\hline
\hline
amplitiude of $e_{A_1A_2A_3A_4}$&1&1&1&1&1&1&1&1&-1&-1&1&1&1&1&-1&-1\\
\hline
\end{tabular}\end{small}
\caption{Traces of matrices}\label{trace1to1}
\end{center}
\end{table}

\noindent
which give us the following multivector:
\begin{eqnarray}
F_n U_f E_n e_{0000}&=&e_{0000}+e_{0001}+e_{0010}+e_{0011}+e_{0100}+e_{0101}+e_{0110}+e_{0111}\nonumber \\
										& &-e_{1000}-e_{1001}+e_{1010}+e_{1011}+e_{1100}+e_{1101}-e_{1110}-e_{1111}
\end{eqnarray}
Note that it is the same as (\ref{wynik121}).

Let us consider the function with non-trivial mask $s=(10)$:\\
\begin{table}[h]
\begin{center}
\begin{tabular}{|c|cccc|}
\hline
\multirow{2}{*}{$X$}& 0&0&1&1\\
    & 0&1&0&1\\
\hline \hline
\multirow{2}{*}{$f(X)$}& 1&0&1&0\\
			 & 0&1&0&1\\
\hline
\end{tabular}\caption{Function with mask $s=(10)$}
\end{center}\end{table}\\
For some technical reason it is less dimensional function than (Table \ref{fmask}) from the previous section.\\Note that $e_{0000}$, $E_n=E_n e_{0000}$ and $F_n$ are exactly the same as (\ref{e0000}), (\ref{Ene0000}) and (\ref{Fn}) respectively.
We have
\begin{equation}
\begin{tiny}
U_f E_n e_{0000}=
\left( \begin{array}{cccccccccccccccc}
0&1&-1&0&0&0&0&0&0&0&0&0&1&0&0&-1\\
-1&0&0&1&0&0&0&0&0&0&0&0&0&1&-1&0\\
1&0&0&-1&0&0&0&0&0&0&0&0&0&1&-1&0\\
0&-1&1&0&0&0&0&0&0&0&0&0&1&0&0&-1\\
0&0&0&0&0&1&-1&0&1&0&0&-1&0&0&0&0\\
0&0&0&0&-1&0&0&1&0&1&-1&0&0&0&0&0\\
0&0&0&0&1&0&0&-1&0&1&-1&0&0&0&0&0\\
0&0&0&0&0&-1&1&0&1&0&0&-1&0&0&0&0\\
0&0&0&0&1&0&0&-1&0&1&-1&0&0&0&0&0\\
0&0&0&0&0&1&-1&0&-1&0&0&1&0&0&0&0\\
0&0&0&0&0&1&-1&0&1&0&0&-1&0&0&0&0\\
0&0&0&0&1&0&0&-1&0&-1&1&0&0&0&0&0\\
1&0&0&-1&0&0&0&0&0&0&0&0&0&1&-1&0\\
0&1&-1&0&0&0&0&0&0&0&0&0&-1&0&0&1\\
0&1&-1&0&0&0&0&0&0&0&0&0&1&0&0&-1\\
1&0&0&-1&0&0&0&0&0&0&0&0&0&-1&1&0\\
\end{array}\right)\end{tiny}
\end{equation}
and
\begin{equation}
\begin{tiny}
F_n U_f E_n e_{0000}=
\left( \begin{array}{cccccccccccccccc}
-2i&2&-2&2i&0&0&0&0&0&0&0&0&2&2i&-2i&-2\\
-2&2i&-2i&2&0&0&0&0&0&0&0&0&2i&2&-2&-2i\\
2&-2i&2i&-2&0&0&0&0&0&0&0&0&2i&2&-2&-2i\\
2i&-2&2&-2i&0&0&0&0&0&0&0&0&2&2i&-2i&-2\\
0&0&0&0&-2i&2&-2&2i&2&2i&-2i&-2&0&0&0&0\\
0&0&0&0&-2&2i&-2i&2&2i&2&-2&-2i&0&0&0&0\\
0&0&0&0&2&-2i&2i&-2&2i&2&-2&-2i&0&0&0&0\\
0&0&0&0&2i&-2&2&-2i&2&2i&-2i&-2&0&0&0&0\\
0&0&0&0&2&2i&-2i&-2&-2i&2&-2&2i&0&0&0&0\\
0&0&0&0&2i&2&-2&-2i&-2&2i&-2i&2&0&0&0&0\\
0&0&0&0&2i&2&-2&-2i&2&-2i&2i&-2&0&0&0&0\\
0&0&0&0&2&2i&-2i&-2&2i&-2&2&-2i&0&0&0&0\\
2&2i&-2i&-2&0&0&0&0&0&0&0&0&-2i&2&-2&2i\\
2i&2&-2&-2i&0&0&0&0&0&0&0&0&-2&2i&-2i&2\\
2i&2&-2&-2i&0&0&0&0&0&0&0&0&2&-2i&2i&-2\\
2&2i&-2i&-2&0&0&0&0&0&0&0&0&2i&-2&2&-2i\\
\end{array}\right)\end{tiny}
\end{equation}
Again using (\ref{ca2mu}) we can find the amplitudes (Table \ref{tracefmask})

\begin{table}[h]
\begin{center}\begin{small}
\begin{tabular}{|c|cccccccccccccccc|}
\hline
$A_1$&0&0&0&0&0&0&0&0&1&1&1&1&1&1&1&1\\
$A_2$&0&0&0&0&1&1&1&1&0&0&0&0&1&1&1&1\\
$A_3$&0&0&1&1&0&0&1&1&0&0&1&1&0&0&1&1\\
$A_4$&0&1&0&1&0&1&0&1&0&1&0&1&0&1&0&1\\
\hline
\hline
Tr($e_{A_1A_2A_3A_4} F_n U_f E_n e_{0000}$)&0&32&32&0&0&-32&-32&0&0&32&-32&0&0&-32&32&0\\
\hline
\hline
amplitiude of $e_{A_1A_2A_3A_4}$&0&2&2&0&0&2&2&0&0&-2&2&0&0&2&-2&0\\
\hline
\end{tabular}\end{small}
\caption{Traces of matrices}\label{tracefmask}
\end{center}
\end{table}

\noindent
which give us the following multivector:
\begin{equation}
F_n U_f E_n e_{0000}=2(e_{0001}+e_{0010}+e_{0101}+e_{0110}-e_{1001}+e_{1010}+e_{1101}-e_{1110})
\end{equation}
Performing (\ref{jedynki}-\ref{wynik}) as we did in Sec. \ref{examples} using the function (Table \ref{fmask}) we find the mask $s=(10)$.

\section{Acknowledgement}

We are deeply indebted to Marek Czachor for stimulating sugestions, help and encouragement.

\end{document}